\documentclass[aps,nofootinbib]{revtex4}
\usepackage[mathscr]{eucal}
\usepackage[dvips]{graphicx}
\usepackage{amsfonts,amssymb,amsthm}
\usepackage{amsmath}
\usepackage{color}

\newcommand{\Tr}{\mathop{\mathrm Tr}\nolimits}

\begin{document}
\title{Emergent matter from 3d generalised group field theories}
%\title{From group field theory to simplicial gravity path integrals through non-commutative geometry}

%\date{December 18,
%2001}
\author{{\bf Alessandro Di Mare} and {\bf Daniele Oriti} \\ Albert Einstein Institute \\ Am Muehlenberg 1, D-14476 Golm, \\ Germany, EU}
\begin{abstract}
We identify classical solutions of a generalised group field theory model in
3 dimensions, and study the corresponding perturbations, deriving their
effective dynamics. We discuss their interpretation as emergent matter
fields. This allows us, on the one hand to test the proposed mechanism for
emergence of matter as a phase of group field theory, and on the other hand
to expose some limitations of the generalised group field theory formalism.
\end{abstract}
\maketitle

\section{Introduction}
In the last decades several approaches to quantum gravity have been developed, with important results \cite{quantumgravity}. Among them, Group Field Theories (GFT) \cite{gftdaniele,gftdanieledue,laurentgft} are, in our opinion, particularly promising. GFTs are quantum field theories defined over a group manifold, representing a \lq\lq meta-space\rq\rq of discrete spacetime geometries, and not  spacetime itself, as a new algebraic and combinatorial realization of the \lq\lq third quantization\rq\rq idea \cite{thirdquantization}. Moreover, GFTs bring together most of the ingredients entering in other \textit{non perturbative} and \textit{background independent} approaches (such as loop quantum gravity, spin foam models and simplicial quantum gravity approaches) \cite{gftdaniele,gftdanieledue}.
Very little is known, still, about group field theory models for quantum gravity, in both 3 and 4 dimensions, and a lot of technical work should go into their analysis, for example their classical solutions, which will be one focus of the present work, and their purely field theoretic aspects, which are the subject of attention at present \cite{danielerazvan,gftscaling, razvan, gftscaling2}.

An important issue in many (non-perturbative) quantum gravity formalisms is the inclusion and the correct description of matter fields. In the literature we can distiguish two different strategies.
One approach is start from a model describing a quantum spacetime without matter and then add degrees of freedom describing matter, being fields, point particles or extended objects. This is the standard route both in loop quantum gravity \cite{thomas,carlo}, spin foam models \cite{PR1,PR3, ioTim, danielehendryk, simoneYM} and simplicial quantum gravity \cite{hamber, DTmatter}. In a GFT context this means, for example, writing down a coupled GFT action  \cite{particle} for both gravity and matter fields that would produce, in perturbative expansion, a sum over simplicial complexes with dynamical geometry together with Feynman graphs for the matter fields living on the
simplicial complexes. Or it means \cite{richard} defining a GFT model whose Feynman amplitudes can be understood as state sum models for gravity with appropriate matter field observables. An alternative approach is to think of matter degrees of freedom as a subset of the same degrees of freedom defining the microstructure of quantum spacetime itself, and emerge as fluctuations around a background configuration of the same. This idea, pursued for example (in different ways) in \cite{sundance, fotini, etera, florian}, shares some similarities with analog gravity models in condensed matter systems, in which quasi-particles with an effective description as matter fields propagating on curved geometries emerge as fluctuations around stable vacua of the underlying many-body system (e.g. a Bose condensate), but are of course collective excitations of the same microscopic degrees of freedom just as well. A strategy for emergent matter within the GFT framework, in particular,  first studied in \cite{etera, florian, emergent}, is based on the following idea: starting from a fundamental GFT action, one looks for solutions of the corresponding equation of motion; then, one considers perturbations around these solutions, obtaining an effective dynamics for the perturbation field. The task is to identify a class of solutions and of perturbations, such that the effective dynamics for these perturbations takes the form of a matter field theory on an effective spacetime. 

\noindent It turns out \cite{emergent} that the effective matter field theories emerging from GFTs following this procedure are noncommutative quantum field theories on non-commutative spacetimes with a Lie algebra structure, and with a curved momentum (group) manifold. This only makes this strategy all the more interesting. In fact, non-commutative geometry has been often advocated as the appropriate language to describe a quantum spacetime, i.e. a regime which, already far from the full non-perturbative quantum gravity dynamics at the (ultra-)Planckian scales, is still semi-classical in that it incorporates somehow the quantum gravity corrections to continuum physics at the effective level. This can take the form of an effective minimal (Planck) length scale in matter field theories, of a non trivial commutation relation between position operators in quantum mechanics (so that the notion of spacetime point, and of spacetime continuum, becomes meaningless), or of a generalised uncertainty principle, or of a form of co-gravity, i.e. curvature in momentum space \cite{ncft,majid}. It is actually this sort of effect, a curved momentum sector for particle (and field) kinematics, which shows up most naturally in a GFT context. Moreover,  in recent years much work in the context of quantum gravity phenomenology \cite{phenomenology} was developed. On the one hand, this starts from the simple idea that several astrophysical systems could work as a magnifying lens to amplify quantum gravity effects even if Planck scale suppressed, and make them accessible to experiments. On the other hand, much of this work relies on effective models of quantum gravity, rather than on the tentative fundamental formalisms currently available, and in particular on effective non-commutative models for matter kinematics and dynamics, one important example being so-called Deformed (or Doubly) Special Relativity \cite{dsr}. This means that the recent results on the emergence of effective non-commutative matter field theories from GFT have the potential to help bridging the gap between the microscopic dynamics they define for quantum space and macroscopic physics and quantum gravity phenomenology.
 
 \
 
\noindent In this paper we thus follow this strategy, and apply it to a generalised class of GFT models introduced in \cite{new,new2,new3}, focusing on the 3d case. This generalised GFT formalism has been introduced as an attempt to formulate GFT models, which on the one hand depended explicitly on metric-related variables, thus allowing a more manifest encoding of simplicial geometry at the level of the GFT action, and on the other hand possessed Feynman amplitudes with the explicit form of a simplicial gravity path integral, thus clarifying the relation between spin foam and simplicial gravity structures. Obviously, the viability and correctness of such a generalised GFT formalism is something to be tested. Therefore, we understand the results of this paper in two dual ways: 1) as a test of the strategy for the emergence matter from group field theories as perturbations around classical GFT solutions; in fact, we check here whether this strategy works nicely also for (much) less trivial models than those it has been applied to so far, and under which assumptions and conditions; 2) as a test of the generalised GFT formalism itself; in fact, we will see that some difficulties encountered in applying this procedure, and some unsatisfactory features of the resulting effective actions for perturbations, can naturally be understood as stemming from limitations of the generalised GFT formalism itself, rather than as failures of the \lq\lq emergent matter strategy\rq\rq. Let us also stress that the same identification of classical solutions of the model and of the corresponding perturbative dynamics will be an interesting and highly non-trivial result, from a purely technical perspective.

\

\noindent The outline of the paper is as follows. In the remaining sections of the introduction, we recall how a non-commutative matter field theory for a scalar field emerges from the perturbations around classical GFT solutions in a simple 3d model, and then we introduce the 3d generalised GFT model. In section II, we consider a simplified version of the latter, identify a class of solutions of the classical GFT equations, and then study the corresponding perturbations; we will see that the basic elements of the complete analysis of perturbations to be done on the full model are already evident, together with the limitations of the same model in this respect. In section III, we move on to the complete model; we find some exact and approximate classical solutions of its equations of motion, and extract the effective dynamics for the corresponding perturbations, focusing on the approximate case. We conclude with a discussion of the results obtained and an outlook on further developments, as suggested  also by our results. 

\subsection{Non-commutative matter from GFT}

In this section we briefly review some results about the emergence of an effective noncommutative field theory in the 3d case \cite{etera}. We start from the following GFT action (Boulatov model), based on a group manifold being either  $SO(3)$ or $SU(2)$:

\begin{equation}
S[\phi]=\frac{1}{2}\int dg_1 dg_2 dg_3\ \phi(g_1, g_2, g_3) \phi(g_3, g_2, g_1)
-\frac{\lambda}{4!}\int dg_1\ldots dg_6\ \phi(g_1,g_2,g_3) \phi(g_3,g_4,g_5)\phi(g_5,g_2,g_6)\phi(g_6,g_4,g_1),
\end{equation}
\\
where $\phi$ is a real field, that we suppose invariant under diagonal action of the group, $\phi(g_1, g_2, g_3)=\phi(g_1h, g_2h, g_3h)$. The corresponding classical equation of motion is

\begin{equation}
\phi(g_3, g_2, g_1)= \frac{\lambda}{3!}\int dg_4 dg_5 dg_6\ \phi(g_3,g_4,g_5)\phi(g_5,g_2,g_6)\phi(g_6,g_4,g_1).
\end{equation}
\\
and a class of solutions of the same equation is parametrized by a function $f: G\rightarrow\mathbb{R}$, satisfying $\int_G dg\ f^2(g)=1$, and given by:

\begin{equation}
\phi_f(g_1, g_2, g_3)= \sqrt{\frac{3!}{\lambda}}\int_G dh\ \delta(g_1 h)f(g_2 h)\delta(g_3 h).
\label{eq:soluzionesemplice}
\end{equation}

Since the Boulatov model defines a (3rd) quantization of BF theory, whose only configurations are locally flat geometries on arbitrary topology, such solutions can be interpreted as describing a quantum flat space. Now we want to study classical perturbations around these solutions, in particular those of the form: $\psi(g_1, g_2, g_3)\equiv\psi(g_1, g_3)=\psi(g_1g_3^{-1})$. For them we obtain the following effective action:

\begin{eqnarray}
S^{(f)}[\psi]&\equiv& S[\phi_f+\psi]-S[\phi_f] = \int dg \psi(g) \mathcal{K}_f(g)\psi(g^{-1})-\frac{\mu}{3!}\int dg_1\ dg_2\ dg_3\ \delta(g_1g_2g_3) \ \psi(g_1)\psi(g_2)\psi(g_3)+\nonumber\\
&-& \frac{\lambda}{4!} \int dg_1\ dg_2\ dg_3\ dg_4\ \delta(g_1\ldots g_4)\ \psi(g_1)\psi(g_2)\psi(g_3)\psi(g_4).
\label{eq:minchia}
\end{eqnarray}
\\
with kinetic term given by

\begin{equation}
\mathcal{K}_f(g)=\frac{1}{2}\left[1-2\left(\int dh\ f(h)\right)^2-\int dh\ f(h) f(hg)\right].
\end{equation}
\\

\noindent The nature of this action as an effective scalar field theory on a noncommutative space is evident once we interpret the group manifold on which it is defined as momentum space, so that the constraints $\delta(g_1g_2g_3)$ and $\delta(g_1\ldots g_4)$ represent the momentum conservation. As a curved space, this momentum space is dual to a noncommutative configuration space with a Lie algebra structure (given by the $\mathfrak{su}(2)$). This duality can be made explicit by introducing a Fourier transform mapping functions over the group to functions over the corresponding algebra \cite{ncgft,laurentmajid}, using which we can rewrite the above action in term of functions on $\mathbb{R}^3$ endowed with a noncommutative $\star$-product \cite{ncgft, laurentmajid}, which can then be shown to be invariant under the quantum double of $SU(2)$, a deformation of the euclidean 3d Poincar\'e group.

\

\noindent Similar results have been obtained also in the 4d case and in Lorentzian signature \cite{florian}, where it has been shown that, starting from a GFT model for $SO(4,1)$ BF theory, one can obtain an effective non-commutative field theory of the DSR type, i.e. for a scalar field living on the $\kappa$-Minkowski spacetime and with a De Sitter space of momenta. Obviously, this is a model of a more direct interest for quantum gravity phenomenology.

\noindent We can conclude that perturbations around a non trivial quantum geometric background, defined by a solution of the GFT equations, follow the dynamics of an effective scalar field theory of a noncommutative type. So we can interpret matter as a phase of quantum geometry, therefore already contained in our quantum gravity models. 
Moreover, this type of analysis of GFT perturbations is useful to expose the deep link between the GFT formalism and noncommutative geometry.

\subsection{Generalised GFT model}
A generalised GFT formalism has been introduced in \cite{new} and further developed in both 3d and 4d in \cite{new2,new3}, its purpose being encoding and controlling in a more explicit way simplicial geometry at the level of the GFT action, and obtaining Feynman amplitude with the manifest form of a simplicial gravity path integral. In turn, this should serve as a complete definition of a quantum dynamics of simplicial structures, to obtain a spin foam model with a clear underlying geometry and  to strengthen the links between spin foam and simplicial gravity approaches.  

In the 3d case, the fundamental  variable is a real function of $3$ group variables and $3$ algebra variables: $\phi:G^3\times\mathfrak{g}^3 \rightarrow\mathbb{R}$,  which satisfies $\phi(g_1,g_2,g_3; B_1,..,B_3)=\phi(g_1h, g_2h,g_3h; hB_1h^{-1},hB_2h^{-1},hB_3h^{-1})$. The group and Lie algebra variables are interpreted as the discrete triad and connection variables of simplicial BF theory or, equivalently, of discrete 3d gravity. 
Its classical action is the following:

\begin{eqnarray}
S&=&\frac{1}{2}\frac{1}{(2\pi)^9}\int_{G^3} dg_1 dg_2 dg_3\int_{\mathfrak{g}^3} d\vec{B}_1 d\vec{B}_2 d\vec{B}_3 \ \phi_{123}\hat{O}_1\hat{O}_2\hat{O}_3\phi^*_{123}+\nonumber\\
&-&\frac{\lambda}{4!(2\pi)^{36}}\int_{G^6} dg_1...dg_6\int_{\mathfrak{g}^6}d\vec{B}_1...d\vec{B}_6 \ \phi_{123}\phi_{345}\phi_{526}\phi_{641},
\label{eq:azione}
\end{eqnarray}
\\
where we have introduced the operator $\hat{O}=\square_G+B^2+\frac{m^2}{8}$, with $\square_G$ the Laplace-Beltrami operator on the group manifold $G$, and the notation $\phi_{123}=\phi(g_1,g_2,g_3; B_1,B_2,B_3)$.

Notice that, even though the field is a function of Lie algebra triad variables, it is treated in this formalism as an ordinary function, and the Lie algebra itself is considered in its vector space aspects only, so that integrations are performed by means of the Lebesgue measure and no star product is introduced when multiplying more than one GFT field. This will have important consequences when studying the corresponding perturbations around classical solutions.

\

For the achievements and limitations of this formalism concerning its original goals, mentioned above, we refer to the literature. Here we will only try to see whether the results on perturbations obtained for the Boulatov model extend to this more complicated case, and at the same time use this analysis of perturbations as a test for the formalism, trying to expose further its limitations by making them explicit at the level of the resulting effective dynamics.
\section{The static-ultralocal case}
Let us begin from the analysis of the simpler \lq\lq static-ultralocal\rq\rq truncation of the model. It corresponds to the case in which the kinetic operator is "freezed" to the identity, imposing $\hat{O}=\mathbb{I}$ in (~\ref{eq:azione}):

\begin{eqnarray}
S=\frac{1}{2}\frac{1}{(2\pi)^9}\int_{G^3} dg_1 dg_2 dg_3\int_{\mathfrak{g}^3} d\vec{B}_1 d\vec{B}_2 d\vec{B}_3 \ \phi_{123}\phi_{321}
-\frac{\lambda}{4!(2\pi)^{36}}\int_{G^6} dg_1...dg_6\int_{\mathfrak{g}^6}d\vec{B}_1...d\vec{B}_6 \ \phi_{123}\phi_{345}\phi_{526}\phi_{641}.
\end{eqnarray}
\\

In the following we \textit{always} assume that \textit{solutions depend on the moduli of the vectors $B$}, and not from its direction, so that the invariance property of the field under the diagonal group action on the Lie algebra arguments $hBh^{-1}$ is trivialized. Moreover, to semplify the notation, we define $B=\left|\vec{B}\right|$. Anyway we use the notation $\int d\vec{B}$, to distinguish integration over all algebra with integration over only the module $\int dB$.

\subsection{Solutions of the equations of motion}
\noindent The equation of motion corresponding to the above action is:

\begin{equation}
\phi_{123}=\frac{\lambda}{3!}\frac{1}{(2\pi)^{27}}\int_{G^3} dg_4 dg_5 dg_6\int_{\mathfrak{g}^3}d\vec{B}_4 d\vec{B}_5 d\vec{B}_6\ \phi_{345}\phi_{526}\phi_{641}.
\end{equation}
\\
and it is easy verify that the following class of functions, a simple generalization of (~\ref{eq:soluzionesemplice}), define classical solutions:

\begin{equation}
\varphi_{123}=\sqrt{\frac{3!(2\pi)^{27}}{\lambda}}\int_G dh\ \delta(g_1h)\psi_1(B_1)f(g_2h; B_2)\psi_3(B_3)\delta(g_3h),
\end{equation}
\\
where the functions $f$, $\psi_1$ and $\psi_3$ have to fulfil the following relations:

\begin{equation}
\int_G dg \int_{\mathfrak{g}} d\vec{B}\ f^2(g,B)=1 \ \ \ \ \ \int_{\mathfrak{g}}d\vec{B}\ \psi_1(B)\psi_3(B)=1.
\end{equation}
\\
We also require, for simplicity,  the function $f(g,B)$ to be separable in its variables: $f(g,B) = f_1(g)f_2(B)$.

\subsection{Perturbations}
Let us consider a generic perturbation $\epsilon(g_1,g_2,g_3; B_1,B_2,B_3)$, and compute the effective action:
\begin{eqnarray}
S[\epsilon]&\equiv&S[\varphi+\epsilon]-S[\varphi]\,=\,\frac{1}{2}\int [dg]\int [d\vec{B}] \ \epsilon_{123}\epsilon_{321}+\nonumber\\
&-& \int [dg] dh_1\int [d\vec{B}] \psi_1(B_5)
f_1(g_2h_1)f_2(B_2)f_1(g_4h_1)f_2(B_4)\psi_3(B_1)\; \epsilon(h_1^{-1},g_2,g_3;B_1,B_2,B_3)\epsilon(g_3,g_4,h_1^{-1}; B_3,B_4,B_5)+\nonumber\\
&-&\frac{1}{2}\int [dg] dh_1 dh_2\int [d\vec{B}] \psi_1(B_3)
f_1(g_4h_1)f_2(B_4)\psi_3(B_1)\psi_3(B_5)\psi_1(B_6)f_1(g_4h_2)f_2(B_4)\psi_3(B_6)\times\nonumber\\
&\times & \epsilon(h_2^{-1},g_2,h_1^{-1};B_1,B_2,B_3)\epsilon(h_1^{-1},g_2,h_2^{-1}; B_5,B_2,B_6)-\frac{\lambda}{4!(2\pi)^{27}}\int [dg] \int [d\vec{B}] \ \epsilon_{123}\epsilon_{345}\epsilon_{526}\epsilon_{641}\, +\nonumber\\
&-&\sqrt{\frac{\lambda}{3!(2\pi)^{27}}}\int [dg] dh \int [d\vec{B}] \ \psi_1(B_6)f_1(g_4h)f_2(B_4)\psi_3(B_1)\; \epsilon(h^{-1},g_2,g_3;B_1,B_2,B_3)\epsilon_{345}\epsilon(g_5,g_2,h^{-1}; B_5,B_2,B_6),\nonumber
\end{eqnarray}
\\
where we used the notation $\epsilon_{123}=\epsilon(g_1,g_2,g_3;B_1,B_2,B_3)$, where convenient.

\noindent This is the action for a generic perturbation, only assumed to have the sam diagonal invariance of the fundamental GFT field, but it simplifies considerably, and can then be interpreted physically, by appropriate simplified choices on the perturbation field. In particular, having in mind the possible interpretation of the $B$ and $g$ variables as configuration and momentum variables, respectively, we try to reduce the dependence of the perturbation field to either of the two sectors. 

\noindent \textbf{Case 1: \textit{Dynamical group sector}}
\noindent Now we show how one can recover the same effective action in \cite{etera} with a particular choice of perturbation field. We assume a separable dependence on $g$ and $B$ variables (this is suggested also by the fact that neither the kinetic nor the interaction term couple the two sectors), and the dependence on a single group variable via the combination $g_1g_3^{-1}$: 
$\epsilon(g_1,g_2,g_3;B_1,B_2,B_3)=\phi(g_1g_3^{-1})\xi(B_1,B_2,B_3)$.
Factoring out the integration over the $B$ variables, we get the following effective action for the perturbation field $\phi(g)$:

\begin{eqnarray}
S[\phi]=\int dg\ \phi(g) \mathcal{K}(g)\phi(g^{-1})
-\frac{\eta}{3!}\int [dg]^3\ \delta(g_1\ldots g_3) \psi(g_1)\ldots\psi(g_3)
- \frac{\mu}{4!} \int [dg]\ \delta(g_1\ldots g_4) \psi(g_1)\ldots\psi(g_4).
\end{eqnarray}
\\
\noindent Here, the kinetic term is given by

\begin{equation}
\mathcal{K}_f(g)=\frac{1}{2}\left[\alpha-2\beta\left(\int dh\ f(h)\right)^2-\gamma\int dh\ f(h) f(hg)\right],
\end{equation}
\\
which is the same of the action (~\ref{eq:minchia}), with up to some appropriate redefinition of coupling constants:

\begin{eqnarray}
\alpha &=& \int d\vec{B}_1 d\vec{B}_2 d\vec{B}_3\ \xi(B_1,B_2,B_3)\xi(B_3,B_2,B_1),\nonumber\\
\beta &=& \int d\vec{B}_1\ldots d\vec{B}_5\ \psi_1(B_5)f_2(B_2)f_2(B_4)\xi(B_1,B_2,B_3)\xi(B_3,B_4,B_5),\nonumber\\
\gamma &=& \int d\vec{B}_1\ldots d\vec{B}_6\ \psi_1(B_3)f_2(B_4)\psi_3(B_5)\psi_1(B_6)f_2(B_4)\psi_3(B_6) \xi(B_1,B_2,B_3)\xi(B_3,B_4,B_5),\nonumber\\
\eta &=&\sqrt{6\lambda}\int dg f_1(g) \int d\vec{B}_1\ldots d\vec{B}_6\ \psi_1(B_6)\psi_3(B_1)f_2(B_4)
\xi(B_1,B_2,B_3)\xi(B_3,B_4,B_5)\xi(B_5,B_2,B_6)\nonumber\\
\mu &=& \int d\vec{B}_1\ldots d\vec{B}_6\ \xi(B_1,B_2,B_3)\xi(B_3,B_4,B_5)\xi(B_5,B_2,B_6)\xi(B_6,B_4,B_1).\nonumber
\end{eqnarray}
\\
\noindent Of course the perturbation field must be chosen such that this new set of coupling constants does not diverge, with the functional dependence on the B sector to be considered as fixed, with no variation allowed in the classical theory and no fluctuations (functional integration) in the quantum theory. We see that we have recovered, even from the more complicated GFT formalism, the expected effective action for the perturbations, that can be interpreted as a noncommutative scalar field theory written in momentum space. Even in this drastic simplification of the formalism, we also see, however, that the way we could obtain it was to completely decouple the B and g sector and to leave only this last one as dynamical. 

\noindent \textbf{Case 2: \textit{Dynamical Lie algebra sector}}
\noindent Now we focus instead on the $B$ sector, and treat the group sector as non-dynamical. The idea is then to reproduce the same action as above, or a modification thereof, but written now in configuration space, according to the standard interpretation of the effective theory and of the variables appearing in the GFT, outlined above. 

\

\noindent As a regularization, we introduce a cut-off $B_{\text{max}}$ in the B integration, and the constant $V_B= 4\pi\int_0^{B_{\text{max}}} dB B^2$, that is the (truncated) volume of $\mathbb{R}^3$. We then consider a perturbation of the form

\begin{equation}
\epsilon_{123}= \frac{1}{\sqrt{V_B}}\,\delta(\vec{B}_1-\vec{B}_2)\;\psi(\vec{B}_2+\vec{B}_3)
\end{equation}
with no dependence on group variables (notice that this does not lead to any divergence because of the compact nature of the group manifold). The dependence on the three variables $B_i$ has been chosen to be what would have followed from a Fourier transform from $g$ to $B$ variables of the function we used as perturbation in the previous case. Moreover we introduce a new condition on the solution imposing that $\int dg f(g)=0$. With this condition the integration over the group variables kills the cubic interaction and some quadratic terms. The resulting effective action for this perturbation field is:

\begin{equation}
S[\psi]= \frac{1}{2} \int d\vec{B}\ \psi^2(B) + \frac{\lambda}{4!}\int d\vec{B}\ \psi^4(B)
\end{equation}

that is an ordinary local $\lambda\phi^4$ theory with a trivial kinetic term (i.e. again a static-ultralocal field theory). 

\

\noindent We can interpret this as a scalar field theory written in configuration space (given by $\mathbb{R}^3$). The triviality of the kinetic term arises directly from the initial ultra-locality restriction, but could possibly be lifted by some more involved choice of classical solution or perturbation field, at the cost of introducing derivatives in the same. The point to notice, however, is that we find no sign of the expected underlying non-commutativity in this field theory. That is, while we find the expected local dependence on $B$ variables that matches their interpretation as configuration variables in some sense dual to the group variables, and while we could try to modify our construction to obtain a more complicated (and interesting) kinetic term or interaction, it is evident that the conjugate nature of $B$ and $g$ sectors is lost and cannot be recovered. Due to the simplicity of the underlying model, it is also clear that this conjugate nature fails to be present in the effective action because it fails to be correctly implemented in the original generalised GFT formalism. We will come back to this point in the following.   

%%%%%%%%%%%%%%%%%%%%%%%%%%%%%%%%%%%%%%%%%%%%%%%%%%%%%%%%%%%%%%%%%%%%%%%%%%%%%%%%%%%%%%%%%%%%%%%%%%%%%%%%%%%%%%%%%%

\section{Full generalised model}
\noindent In this section we consider the complete action (~\ref{eq:azione}) and the corresponding integro-differential equation of motion :
\begin{eqnarray}
\hat{O}_1\hat{O}_2\hat{O}_3\phi_{321}
=\frac{\lambda}{3!(2\pi)^{27}}\int_{G^3} dg_4 dg_5 dg_6\int_{\mathfrak{g}^3}d\vec{B}_4 d\vec{B}_5 d\vec{B}_6 \ \phi_{345}\phi_{526}\phi_{641}.
\label{eq:difficile}
\end{eqnarray}

\subsection{Exact solution of the equation and perturbations}
We first try to find an exact solution of these equation and to obtain an effective field theory for the perturbations.
We look for solutions of the type:
\begin{equation}
\phi_{123}=\int dh\ \psi_1(g_1h,B_1)f(g_2h,B_2)\psi(g_3h,B_3),
\label{eq:soluzione}
\end{equation}
\\
where we have imposed the usual invariance under diagonal right action of the group, by explicit projection.  

It is then easy to show that any function $\phi_{123}$ of the above form is an exact solution of the GFT equation of motion if the following conditions are satisfied:

\begin{enumerate}
	\item the functions $\psi_1, f, \psi_3$ are solutions of the equation
	
	\begin{equation}
	\left(\square+B^2\right)\phi(g,B)=0,
	\label{eq:kernel}
	\end{equation}
	
	so they are eigenfunctions of the operator $\hat{O}$ with eigenvalue $-\frac{m^2}{8}$.
	\item $\psi_1$, $\psi_3$ and $f$ satisfy
	
	\begin{equation}
	\int d\vec{B} \int dg\ \psi_3(gh_1,B)\psi_1(gh_2,B)=\delta(h_1^{-1}h_2)\hspace{1cm}
	\int d\vec{B} \int dg\ f^2(g,B)=1.
	\end{equation}
\end{enumerate}

We now look for such functions, starting from condition 1. We now prove the following:

\textbf{Proposition}: \textit{Let $\mathcal{I}$ be the set of functions defined on the Cartesian product $G\times\mathfrak{g}$ and square integrable with respect to the group variable (so that we can apply the Peter Weyl theorem to decompose them in irreducible representations $j$ of G). The most general solution, in the set $\mathcal{I}$, of the equation (~\ref{eq:kernel}) is of the form}

\begin{equation}
\phi(g,B)=\sum_{j,n,m}\tilde{\phi}_j^{nm}(B)D^j_{nm}(g) \;\;\;\text{with}\;\;\;
\left\{
\begin{array}{rl}
& \phi^j_{mn}(B) \;\;\;\text{arbitrary} \;\;\;\forall B:  B^2=j(j+1)\\
\\
& \phi^j_{mn}(B) = 0 \;\;\;\; \text{elsewhere}.
\end{array}
\right.
\label{eq:eigenfunction}
\end{equation}

The proof is straightforward. 
Because of Peter-Weyl theorem, every solution of the equation, belonging the set $\mathcal{I}$, can be written as $\phi(g,B)=\sum_{j,n,m}\tilde{\phi}_j^{nm}(B)D^j_{nm}(g)$, for some coefficient functions $\phi^j_{mn}(B)$. Being a solution, it has to satisfy: $\sum_{j=0}^{\infty}\sum_{n,m}\left\{\left[-j(j+1)+B^2\right]\tilde{\phi}_j^{nm}(B)\right\}D^j_{nm}(g)=0$, which implies, since the $D^{j}_{mn}(g)$ is a complete orthonormal basis in $\mathcal{I}$, that:  $[-j(j+1)+B^2]\tilde{\phi}_j^{nm}(B)=0 \quad \forall j,n,m$. This equation can be satisfied only if the coefficient functions $\phi^j_{mn}(B)$ are null for all values such that $B^2 \neq j(j+1)$, while they can be arbitrary otherwise.
\hfill$\square$

\

Therefore, the functions (~\ref{eq:eigenfunction}), considered as function of $\vec{B}$ are non null only on spheres of radius $\sqrt{j(j+1)}$. Thus the integral over the algebra of these eigenfunctions is obviously zero, because each sphere is a set of null measure. Therefore the conditions 2 and 3 unless the functions $\tilde{\phi}_j^{nm}(B)$ are actually distributions on $\mathbb{R}^3$. The idea is then to define a 3 dimensional delta-like function, i.e. null everywhere, except on the origin and on the spheres of radius $\sqrt{j(j+1)}$, with $j\in\mathbb{N^{+}}$, where it goes to infinity, and multiply it by a regular function of $B$ to define completely the functions $\tilde{\phi}_j^{nm}(B)$ . There are many possibilities to define this distribution with a limit procedure, and we do it in appendix, providing explicit examples.
In the following we call that distribution $\alpha(\vec{B})$. Its only property we need is:

\begin{equation}
\int d\vec{B}\ \alpha^2(\vec{B})f(B)=\sum_{j=0}^{\infty}(2j+1)^2 f(\sqrt{j(j+1))}),
\end{equation}
\\
where both integral and sum have to converge. Let $\psi_1(g,B)$ and $\psi_3(g,B)$ solutions of (~\ref{eq:kernel}). We also introduce the following distributions:

\begin{eqnarray}
\varphi_1(g,B)=\alpha(\vec{B})\psi_1(g,B)\hspace{1cm}
\varphi_3(g,B)=\alpha(\vec{B})\psi_3(g,B).
\end{eqnarray}
\\
These define solutions too, as they satisfy condition 1, as it can be easily checked. Now we consider condition 2, and compute the following integral:

\begin{eqnarray}
&&\int dg\ \int d\vec{B} \varphi_1(gh_1,B)\varphi_3(gh_2,B)
=\int dg\ \int dB \alpha^2(\vec{B})\psi_1(gh_1,B)\psi_3(gh_2,B)=\nonumber\\
&=&\sum_{j=0}^{\infty}(2j+1)^2 \int dg\ \psi_1\left(gh_1,\sqrt{j(j+1)}\right)\psi_3\left(gh_2,\sqrt{j(j+1)}\right)\,=\,\delta(h_1h_2^{-1}).\nonumber
\end{eqnarray}
\\
Using the above theorem, we know that

\begin{equation}
\psi\left(g,\sqrt{j(j+1)}\right)=\sum_{nm}\left(\tilde{\psi}(\sqrt{j(j+1)})\right)_j^{nm}D^j_{nm}(g).
\end{equation}
which results in the condition:
\begin{equation}
\left(\tilde{\psi}_1\right)_j^{nm_1}\left(\tilde{\psi}_3\right)_j^{nm_3}=\delta^{m_1m_3} \quad \forall j\in\mathbb{N}\, , 
\end{equation}
which characterizes functions satisfying condition 2.

Using the same distribution $\alpha(\vec{B})$, we can define  a function satisfying condition 3 too. Let $f$ be a function of the type ~\ref{eq:eigenfunction}; we define the distribution
$F(g,B)=\alpha(\vec{B})f(g,B)$, and impose condition 3, to find

\begin{eqnarray}
\int dg\ \int d\vec{B}\ F^2(g,B)=\int dg \int d\vec{B} \alpha^2(\vec{B}) f^2(g,B)=
\ \sum_{j=0}^{\infty}(2j+1)^2\int dg\ f^2(g,\sqrt{j(j+1)}).
\end{eqnarray}
\\
If this sum is convergent then we can normalize $f$ so that the result is 1 and condition 3 is satisfied.

\noindent Therefore, using functions satisfying the above conditons, we have found an exact solution of equation  (~\ref{eq:difficile}):

\begin{eqnarray}
\phi_{123}=\int dh\ \varphi_1(g_1h,B_1)F(g_2h,B_2)\varphi(g_3h,B_3)=
\ \int dh\ \alpha(\vec{B_1})\psi_1(g_1h,B_1)\alpha(\vec{B_2})f(g_2h,B_2)\alpha(\vec{B_3})\psi_3(g_3h,B_3).\nonumber
\end{eqnarray}

Now we write down the effective action for a generic perturbation around such solution:
\begin{eqnarray}
S[\epsilon]&=&\frac{1}{2}\frac{1}{(2\pi)^9}\int [dg]\int [d\vec{B}]\, \epsilon_{123}\hat{O}_1\hat{O}_2\hat{O}_3\epsilon_{321}-\frac{\lambda}{4!(2\pi)^{36}}\int [dg] \int [d\vec{B}] \epsilon_{123}\epsilon_{345}\epsilon_{526}\epsilon_{641}+\nonumber\\
&-& \frac{1}{(2\pi)^9}\int [dg] dh_1 dh_2\int [d\vec{B}]\,\psi_1(g_1h_1,B_1)f(g_2h_1,B_2)f(g_4h_2,B_4)\psi_3(g_5h_2,B_5)\Psi(h_1^{-1}h_2)
 \epsilon_{526}\epsilon_{641}\nonumber\\
&-&\frac{1}{2}\frac{1}{(2\pi)^9}\int [dg] dh_1 dh_2\int [d\vec{B}] \;\psi_1(g_1h_1,B_1)\psi_3(g_3h_1,B_3)\psi_1(g_5h_2,B_5)\psi_3(g_6h_2,B_6)\xi(h_1^{-1}h_2)
 \epsilon_{345}\epsilon_{641}\nonumber\\
&-&\sqrt{\frac{\lambda}{3!(2\pi)^{45}}}\int [dg]\int [d\vec{B}] \psi_1(g_1h, B_1)
 f(g_2h, B_2)\psi_3(g_3h, B_3) \epsilon_{345}\epsilon_{526}\epsilon_{641}\,\,\hspace{0.5cm}
\label{eq:long}
\end{eqnarray}
\\
where we have defined:
\begin{eqnarray}
\Psi(h_1^{-1}h_2)\equiv\int dg \int d\vec{B}\ \psi_3(gh_1,B)\psi_1(gh_2,B)\hspace{1cm}
\xi(h_1^{-1}h_3)\equiv\int dg \int d\vec{B}\ f(gh_1,B)f(gh_3,B).
\label{eq:spacchio}
\end{eqnarray} 

\noindent This is a very general result, holding for any solution and any perturbation. Now we would obtain, starting from ~\ref{eq:long}, an action with a kinetic term of the type $\int dx\ \phi(x)K(x)\phi(x)$, either in group or in Lie algebra variables, imposing some particular form of the perturbation field. This turns out to be very difficult, and we have not been able to identify any simple form of the perturbations that would give such simplification. Roughly speaking, it is due to the presence of the terms $\epsilon_{526}\epsilon_{641}$ and $\epsilon_{345}\epsilon_{641}$, which are not equivalent, because we have not assumed any invariance of the GFT field, nor of the corresponding perturbation, under permutations of its arguments. There is no simple choice of perturbation field that gives a product of the type $\epsilon(x)\epsilon(x)$, for example no field with a dependence on the $B$ variables only through the linear combination $a B_1+ b B_2 + c B_3$, or the analogue in the $G$ variables. This is true, unless we assume that the fundamental GFT field as well as the classical solution chosen and the corresponding perturbation field are invariant under permutations of their group and Lie algebra arguments. Imposing this property however, would complicate the analysis considerably. Leaving aside the issue of perturbations, the problem can be seen as due to the non-separability of the solution chosen, in particular the functions $f$ and $\Psi$.  We cannot exclude that a different solution, or a more involved choice of perturbation would lead to a nicer effective action.  However, we also note that, even if one solves somehow this first problem, another trouble comes from the distribution $\alpha(\vec{B})$, which appears in the effective action for the perturbations with different powers, leading to additional technical complications.

\

\noindent This suggests us to leave aside for the moment the issue of perturbations around exact solutions of the generalised model. We turn our attention, instead, to {\it approximate solutions} and their corresponding perturbations.

\subsection{Approximate solutions and perturbations}
\noindent The basic idea here is to look for eigenfunctions of the Laplace-Beltrami operator and, at the same time, good approximations of the delta function on the group. We note that every good approximation of the delta function on the group has to be sharply peaked on the identity. Therefore the Laplace-Beltrami operator applied to such a function will be close to zero everywhere but in a neighbourhood of the identity element. For this reason we study now the behaviour of the Laplace-Beltrami operator on the group $SU(2)$ which is isomorphic to the 3-sphere $S_3$, very close to the identity element, which we identify with the north pole. Now we fix a chart in a neighborhood of the identity element, and we use polar coordinates. The Laplace-Beltrami operator becomes:

\begin{eqnarray}
\square=\frac{1}{\sqrt{g}}\partial_{\mu}\left(\sqrt{g}g^{\mu\nu}\partial_{\nu}\right)
\approx\frac{1}{\theta^2}\frac{\partial}{\partial \theta}\left(\theta^2\frac{\partial}{\partial\theta}\right)+ \frac{1}{\theta^2\sin\phi}\frac{\partial}{\partial \phi}\left(\sin\phi\frac{\partial}{\partial\phi}\right)+ \frac{1}{\theta^2\sin^2\phi}\frac{\partial^2}{\partial\phi^2}\nonumber
\end{eqnarray}
\\
If we apply this operator to functions $f$ only dependent on the  $\theta$ variable, we can consider only the first term in the previous formula, so we can use the operator:

\begin{equation}
\square_A f=\frac{1}{\theta^2}\frac{\partial }{\partial \theta}\left(\theta^2\frac{\partial f}{\partial\theta}\right)=\frac{1}{\theta}\frac{\partial^2}{\partial \theta^2}\left(\theta f\right),
\end{equation}
\\
where the $A$ indicates both the approximation in the operator and that we can apply it just on function of $\theta$ considerably different from zero only in a small neighbourhood of $\theta=0$.

Now we consider the eigenvalue problem for this operator: $\square_A f(\theta)=\lambda f(\theta)$. Among the solutions of this equation, one is particular interesting for our purposes
$
f(\theta)=\frac{e^{-\frac{\theta}{a}}}{\theta}
$,
where $\lambda=\frac{1}{a^2}$. Indeed we note that in the limit $a\rightarrow 0^+$ the function $f(\theta)$ goes to zero everywhere but in $\theta=0$ where it goes to infinity. Therefore, with appropriate normalization, this is a good representation of the Dirac delta function. Thus we define the function

\begin{equation}
\delta_a(\theta)=\frac{e^{-\frac{\theta}{a}}}{a^2\theta}
\end{equation}
\\
that satisfies the following properties:

\begin{eqnarray}
\square\delta_a(\theta)\,\approx\,\square_A\delta_a(\theta)\,=\,\frac{1}{a^2}\delta_a(\theta)\hspace{1cm}
\lim_{a\rightarrow 0^+}\delta_a(\theta)\,=\,\delta(\theta),
\end{eqnarray}
\\
\noindent Of course we can rewrite the $\delta_a$ function in term of the group variable, using the parametrization: $g=\cos\theta \mathbb{I}+i\sin\theta \hat{n}\cdot\vec{\sigma}$, with $\Tr(g)=2\cos(\theta)$ and $\hat{n}\in S_2$. Through this relation we can interpret the $\delta_a(\theta)$ as a $\delta_a(g)$ on SU(2). Now we apply the operator $\hat{O}$ on $\delta_a(g)$, with $a\approx 0^+$, to find:

\begin{equation}
\left(\square+B^2-\frac{m^2}{8}\right)\delta_a(g)\approx\left(\frac{1}{a^2}+B^2-\frac{m^2}{8}\right)\delta_a(g)\approx\frac{1}{a^2}\delta_a(g).
\label{eq:a}
\end{equation}
\\
Of course the approximation improves as $a$ tends to zero. Indeed we have supposed that $\frac{1}{a^2}>>\left|B^2-\frac{m^2}{8}\right|$. In order for our solution of the equations of motion to satisfy this condition for all relevant values of $B$, we then assume that it is a function with compact support (i.e. identically zero outside of some bounded set) with respect to the $B$ variables. 

\noindent Another important point is that we can handle $\delta_a(g)$ like a true $\delta(g)$ function, as it satisfies $\int dg\ \delta_a(gh^{-1})f(g)\approx f(h)
$.
The last step is to introduce another function $\Delta_a(\theta)$ on the sphere $S_3$. It is defined, for $0\leq\theta<\pi/2$ (i.e. for the the half-sphere corresponding to $SO(3)$), as

\begin{equation}
\Delta_a(\theta)=\frac{1}{2}\frac{1}{\sqrt{\pi a}}\frac{e^{-\frac{\theta}{a}}}{\theta} \,=\frac{1}{2\sqrt{\pi}} a^{\frac{3}{2}}\delta_a(\theta)
\end{equation}
\\
and it is extended to the whole $S_3\approx SU(2)$ (i.e. to $0\leq \theta < \pi$), by requiring $\Delta_a(g)= -\Delta_a(-g)$, again for $a\approx 0$. This function is almost vanishing everywhere but for $g=\mathbb{I}, -\mathbb{I}$ and satisfies:

\begin{equation}
\square\Delta_a(\theta)\approx\frac{1}{a^2}\Delta_a(\theta)\hspace{1cm} \int dg \Delta_a(g)\,=\,0\hspace{1cm}
\int dg\ \Delta^2_a(g)=1.
\end{equation}
\\

\noindent In the last two properties the integration is over $SU(2)$. If we restricted the integration only over $SO(3)$, the first integral gives $2\sqrt{\pi}a^{3/2}$ and the second one gives $1/2$.

\noindent Finally using all the above, it is easy to verify that the following function is an approximate solution of (~\ref{eq:difficile}):

\begin{equation}
\phi_{123}=\sqrt{\frac{3!(2\pi)^{27}}{\lambda a^6}}\int dh\ \delta_a(g_1h)\psi_1(B_1)\Delta_a(g_2h)f(B_2)\delta_a(g_3h)\psi_3(B_3),
\end{equation}
\\
where $\psi_1, f, \psi_3$ are functions with compact support, dependent only on the module of $\vec{B}$, and satisfy the conditions

\begin{equation}
\int d\vec{B}\ f^2(B)=1\ \quad\ \int d\vec{B}\ \psi_1(B)\psi_3(B)=1.
\end{equation}

\

\noindent Next, we study perturbations around this new approximate solutions.
The effective action for a generic perturbation field $\epsilon(g_1, g_2, g_3; B_1, B_2, B_3)$ is:

\begin{eqnarray}
\hspace{-0.8cm} S[\epsilon]&\propto&\frac{1}{2}\int [dg]\int [d\vec{B}] \, \epsilon_{123}\hat{O}_1\hat{O}_2\hat{O}_3\epsilon_{321}\,-\,\frac{\lambda}{4!(2\pi)^{27}}\int [dg] \int [d\vec{B}] \, \epsilon_{123}\epsilon_{345}\epsilon_{526}\epsilon_{641}\,+ \nonumber \\
&-& \frac{1}{a^6}\int [dg] dh\int [d\vec{B}]\, \psi_1(B_5)\Delta_a(g_2h)f(B_2)\Delta_a(g_4h)f(B_4)\psi_3(B_1)\; \epsilon(h^{-1},g_2,g_3;B_1,B_2,B_3)\epsilon(g_3,g_4,h^{-1}; B_3,B_4,B_5)+\nonumber\\
&-&\frac{1}{2}\frac{1}{a^6}\int [dg] dh_1 dh_2\int [d\vec{B}]\, \psi_1(B_3)\Delta_a(g_4h_1)\psi_3(B_5)\psi_1(B_6)\Delta_a(g_4h_2)\psi_3(B_1)\times\nonumber\\
&\times & \epsilon(h_2^{-1},g_2,h_1^{-1};B_1,B_2,B_3)\epsilon(h_1^{-1},g_2,h_2^{-1}; B_5,B_2,B_6)+\nonumber\\
&-&\sqrt{\frac{\lambda}{3!(2\pi)^{27}a^6}}\int [dg] dh\int [d\vec{B}]\, \psi_1(B_6)\Delta_a(g_4h)f(B_4)\psi_3(B_1)\; \epsilon(h^{-1},g_2,g_3;B_1,B_2,B_3)\epsilon_{345}\epsilon(g_5,g_2,h^{-1}; B_5,B_2,B_6)\;\;\;\;\;\;\;
\label{eq:dupalle}
\end{eqnarray}
\\
Here there is a first difference with respect to the previous case: due to the approximation, the linear terms, which would cancel exactly when perturbing around exact extrema of the action, do not cancel exactly but are nevertheless suppressed with $a$ and we have not written them for this reason. 

\noindent In order to get a "good" effective theory we need again to impose some special conditions on the perturbation field. As in the static-ultralocal theory, we have two sets of variables, group and algebra, therefore we have more room for manouvering with respect to previous models. The goal is again to obtain a field theory with a kinetic term written in the usual form $\int dx\ \phi(x)K(x)\phi(x)$, and thus with a field depending on a single (group or algebra) argument. Moreover we recall that the group and algebra variables have the interpretetion of momentum and configuration variables in the usual matter field theory of noncommutative type. 

\noindent The first assumption is the separability of the perturbation field:

\begin{equation}
\epsilon(g_1,g_2,g_3;B_1,B_2,B_3)=\varphi(g_1,g_2,g_3)\phi(B_1,B_2,B_3)
\label{eq:separazione}
\end{equation}

Next, having again in mind the possible interpretation of the $B$ and $g$ variables as configuration and momentum variables, respectively, we try to reduce the dependence of the perturbation field to either of the two sectors, by treating the other sector as non-dynamical, as in the static-ultralocal case. 

\

\noindent \textbf{Case 1: \textit{Dynamical group sector}}
\noindent We recall that we have fixed the value of $a$ very close to zero. Thus we notice immediatly that in the action (~\ref{eq:dupalle}) we have two dominant quadratic terms. We restrict our attention to them, i.e. to the free field theory sector. The dominant terms, comparing powers of $1/a$, are 

\begin{eqnarray}
S[\epsilon]&=&- \frac{1}{(2\pi)^9}\frac{1}{a^6}\int dg_2 dg_3 dg_4 dh_1\int d\vec{B}_1\ldots d\vec{B}_5
\psi_1(B_5)\Delta_a(g_2h_1)f(B_2)\Delta_a(g_4h_1)f(B_4)\psi_3(B_1)\times\nonumber\\
&\times & \epsilon(h_1^{-1},g_2,g_3;B_1,B_2,B_3)\epsilon(g_3,g_4,h_1^{-1}; B_3,B_4,B_5)+\nonumber\\
&-&\frac{1}{2}\frac{1}{(2\pi)^9}\frac{1}{a^6}\int dg_2 dg_4 dh_1 dh_2\int d\vec{B}_1d\vec{B}_2d\vec{B}_3d\vec{B}_5d\vec{B}_6 \psi_1(B_3)\Delta_a(g_4h_1)\psi_3(B_5)\psi_1(B_6)\Delta_a(g_4h_2)\psi_3(B_6)\times\nonumber\\
&\times & \epsilon(h_2^{-1},g_2,h_1^{-1};B_1,B_2,B_3)\epsilon(h_1^{-1},g_2,h_2^{-1}; B_5,B_2,B_6)\nonumber
\end{eqnarray}
\\
Now we separate the variables like in (~\ref{eq:separazione}) and we impose the usual condition $\varphi(g_1,g_2,g_3)=\varphi(g_1g_3^{-1})$. After integrating out $B$ variables, treating the part of the perturbation field depending on them as non-dynamical, we get

\begin{equation}
S[\varphi]=\frac{1}{2}\frac{1}{(2\pi)^9}\frac{C}{a^6}\int dg\ \varphi(g)K(g)\varphi(g^{-1}),
\end{equation}
\\
where the kinetic operator is

\begin{equation}
K(g)=\int dh\ \Delta_a(h)\Delta_a(hg),
\end{equation}
\\
and we have defined the constant

\begin{equation}
C=\int d\vec{B}_1\vec{B}_2\vec{B}_3\vec{B}_5\vec{B}_6 \ \psi_1(B_3)\psi_3(B_5)\psi_1(B_6)\psi_3(B_1)\phi_{123}\phi_{526}.\nonumber
\end{equation}

\noindent We see that we obtain once more a nice field theory on a  single group manifold, interpreted as (curved) momentum space, whose dynamics is again entirely determined by the classical GFT solution chosen, thus by the quantum spacetime we select when selecting the solution itself, and on which the effective scalar field propagates. Once more, we could Fourier transform from group to Lie algebra space, playing the role of effective spacetime, introducing the appropriate star product.

\

\noindent \textbf{Case 2: \textit{Dynamical Lie algebra sector}}
\noindent Here we restrict to the $SU(2)$ case. Without this restriction our choice of perturbation field gives a non-local effective action. This case, from a technical point of view, is very similar to the static-ultralocal limit. Indeed we choose exactly the same type of perturbation field:

\begin{equation}
\epsilon_{123}=\frac{1}{\sqrt{V_B}}\delta(\vec{B}_1-\vec{B}_2)\psi(\vec{B}_2+\vec{B}_3)
\end{equation}

\noindent where $V_B$ is again the (truncated) volume of $\mathbb{R}^3$. After a straightforward calculation,  we get

\begin{equation}
S[\psi]= \frac{1}{128}\int d\vec{B}\ \psi(\vec{B})K(B)\psi(\vec{B}) + \frac{\lambda}{4!} \int d\vec{B}\ \psi^4(\vec{B})
\end{equation}

\noindent where $K(B)=(B^2-\frac{m^2}{2})^3$. This is a nice local effective action with a rather unusual kinematical term, with an algebraic dependence on the configuration space Lie algebra variable. 
Once more, the specific form of the kinetic term could possibly be modified by different choices of classical (approximate) solution or perturbation field, even though it is not straightforward to envisage how to do so. What can not be modified easily is again the fact that we find no sign of the expected underlying non-commutativity in this field theory, and that the conjugate nature of $B$ and $g$ sectors is lost. This confirms that something is dubious in the way the original GFT model, i.e. the generalised GFT formalism, because we can once more trace back the origin of this limitation in the effective field theory to the way the generalised GFT formalism deals with the dependence on the GFT field on the Lie algebra $B$ variables. 

\section{Discussion and conclusions}
\noindent Let us summarize the results we have obtained. We have studied the issue of classical solutions of the generalized 3d GFT model proposed in \cite{new,new2,new3}, and of the effective dynamics of the corresponding perturbations, trying to identify specific solutions and perturbations such that the dynamics of the latter can be understood as that of a scalar field on an effective (noncommutative) spactime. 

\noindent The motivations were the following. First of all, to improve our understanding of such models,  at a more technical level. Second, to show that indeed effective matter field theories could be extracted as phases of such generalised models. This means on the one hand again solving the technical difficulties coming from the more involved structure of this model, as compared to the easier ones which have been dealt with up to now, but on the other hand showing that the idea of matter as emerging from the collective dynamics of the micro-costituents of quantum spacetime is more solid than could be originally thought, holding for a wider class of GFT models. Third, and conversely, to use this procedure for extracting effective matter theory from GFT as a test for the generalised GFT formalism itself. This means learning from the way this strategy for emergent matter is made to succeed or from the way it fails, in this context, what are the weak and dubious points of the formalism. 

\noindent Our results are indeed mixed. We have managed to identify exact solutions of the GFT equations of motion both in the simpler static-ultralocal truncation of the model and in the full, more complicated, one. In the truncated model we could extract rather straightforwardly an effective scalar field theory for the perturbations. When only the group sector of the model is treated as dynamical, the effective field theory takes indeed the expected form of a non-commutative field theory written in momentum space. One could then use a non-commutative Fourier transform to re-write the theory in configuration space, given by the $\mathfrak{su}(2)$ Lie algebra. Because of the presence of both Lie algebra and group variables in the fundamental GFT model, one could also expect that it should be possible to obtain directly a non-commutative  field theory for the perturbations in configuration space, by treating directly the only Lie algebra sector of the perturbations as dynamical. This turns out not to be the case. One obtains indeed an effective scalar field theory on the same Lie algebra, but without the wanted and expected non-commutative structure. This turn out to be the result both in the truncated model and in the full model, for which  we have been able to extract a nice(r) effective dynamics for perturbations, i.e. one that could be interpreted as a scalar field theory on an emergent 3d spacetime, only within a certain approximation, that is only around approximate solutions of the equations of motion. Moreover, a closer look at the way the solutions of the equations of motions could be found and the effective dynamics for perturbations extracted shows that it has been necessary, also when dealing with the full model, to basically trivialize the kinetic term of the theory to the level of the corresponding equations, i.e. look for solutions for which the non-trivial part of the kinetic term gives a basically trivial contribution to the equations. While this last point, although somehow disappointing, could be argued to be solvable with more work and some more clever analysis of the same equations, the first point appears to us to be more fundamental, i.e. to suggest a fundamental limitation of the generalized GFT formalism itself.   

\noindent Let us be more precise. The generalised GFT formalism of \cite{new,new2,new3} is based on the idea of enlarging the set of variables on which the GFT field depends from group elements alone to include Lie algebra elements as well, interpreted as discrete connection and $B$ field of a BF-based description of simplicial geometry, which should become manifest at the level of the Feynman amplitudes. The two sets of variables are treated on equal footing and in parallel, while maintaining the GFT an ordinary, even if  non-local with respect to the combinatorics of arguments, field theory. This implies two choices: first, treating  $\mathfrak{su}(2)$ as  an ordinary vector space, thus neglecting its non-commutative nature, and, second, relaxing  the conjugate nature (at the classical level) of $B$ and $g$ variables in the definition of the model (which characterizes BF theory), and more specifically in the choice of kinetic term. Some consequences of these choices from the point of view of simplicial geometry are discussed in \cite{new2,new3}. 
The results we have presented concerning the effective dynamics of GFT perturbations, in particular the \lq\lq negative\rq\rq ones, suggest that the way the Lie algebra $B$ variables are introduced and treated in the generalised GFT formalism is not correct. The apparent need to trivialize the kinetic term in the choice of classical solution suggests that only when the conjugate nature of the $B$ and $g$ variables is implemented, and thus a BF kynematics is chosen, the corresponding classical solutions define an effective (quantum) flat spacetime as desired. The lack of noncommutative structure in the effective action in the Lie algebra sector suggests instead that the correct way of introducing the $B$ variables in the formalism is not to enlarge the domain space of the GFT field, but  to use from the start a noncommutative Fourier transform (in particular, the one introduced and studied in \cite{PR3, laurentmajid, karim} to go from group space to Lie algebra space, turning the original formulation of GFTs into a noncommutative (and combinatorially non-local) one. All these suggestions are indeed taken into account in the new noncommutative formulation of group field theories developed in \cite{ioaristide}, which, on top of solving several difficulties we encountered here in the study of perturbations, brings many additional  bonuses and further insights from the point of view of simplicial geometry and fundamental quantum gravity proper.

\section*{Appendix: the distribution $\alpha(\vec{B})$}

\begin{figure}[htbp]
\centering
\includegraphics[scale=0.35]{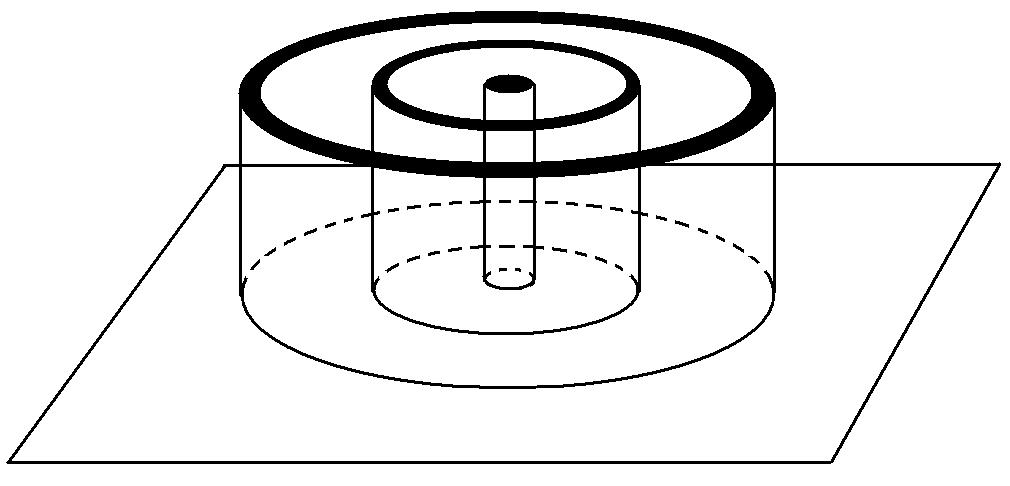} 
\caption{Geometric idea of the distribution $\alpha(\vec{B})$}
\label{fig:cilindri}
\end{figure}

\noindent In this appendix we introduce in a geometrical way the distribution $\alpha(\vec{B})$.
The geometric idea is shown in figure ~\ref{fig:cilindri}, where we have suppressed one dimension. Over each corona, we build a cylindrical one such that each one is built around a 2-sphere of radius $\sqrt{j(j+1)}$, and has thickness $2b$. Just like the Dirac delta function, we then send $b$ to zero keeping the volume of each corona fixed.
Let us define a set of functions:

\begin{equation}
f_j(\vec{B}; b)=
\left\{
\begin{array}{rl}
\frac{(2j+1)^2}{V_b(j)} & \sqrt{j(j+1)}-b\leq\left|\vec{B}\right|\leq \sqrt{j(j+1)}+b\\
\\
0 & elsewhere
\end{array}
\right.
\end{equation}
\\
where b is a positive, but `small', number and $V_b(j)=\frac{4\pi}{3}[2b^3+6j(j+1)b]$ is the 3-volume of the $j^{th}$ spherical corona, for $j\in\mathbb{N}$ and $j>0$. Moreover we define $V_b(0)=\frac{4\pi}{3}b^3$, and

\begin{equation}
f_0(\vec{B}; b)=
\left\{
\begin{array}{rl}
\frac{1}{V_b(0)} & 0\leq\left|\vec{B}\right|\leq b\\
\\
0 & elsewhere
\end{array}
\right.
\end{equation}
\\
Each $f_j$ represents a cylindrical corona, while $f_0$ is the central cylinder. So we have our distribution:

\begin{equation}
\alpha(\vec{B})=\lim_{b\rightarrow0^+}\sqrt{\sum_{j=0}^{\infty}f_j(\vec{B},b)}.
\end{equation}
\\
Of course it is possible to define our distribution through smoother functions, e.g. gaussian-like functions centered in $|\vec{B}|=\sqrt{j(j+1)}$, and then follow the same procedure showed above, with the same result.

\noindent Now, suppose we need to calculate an integral of the type

\begin{equation}
\int d\vec{B}\ \alpha^2(\vec{B})f(|\vec{B}|).
\end{equation}
\\
From a geometric point of view, the height of the $j^{th}$ cylinder, of volume $(2j+1)^2$, is multiplied by $f(\sqrt{j(j+1)})$. Therefore the resulting volume is (with $f$ having spherical symmetry):

\begin{equation}
\int d\vec{B}\ \alpha^2(\vec{B})f(B)=\sum_{j=0}^{\infty}(2j+1)^2 f(\sqrt{j(j+1)}).
\label{eq:distribuzione}
\end{equation}

\end{document}